\documentclass[prl,aps,twocolumn,showpacs]{revtex4}
\usepackage[T1]{fontenc}
\usepackage[latin9]{inputenc}
\usepackage{amsmath}
\usepackage{graphicx}
\usepackage[english]{babel}
\usepackage{bm}
\usepackage{times}

\begin{document}

\title{Reservoirs of Stability: Flux Tubes in the Dynamics of Cortical Circuits}


\author{Michael Monteforte}

\email{monte@nld.ds.mpg.de}

\author{Fred Wolf}

\affiliation{Max Planck Institute for Dynamics and Self-Organization 37073 G\"ottingen,
Germany\\
BCCN, BFNT and Faculty of Physics, University G\"ottingen, 37073
G\"ottingen, Germany}
\begin{abstract}
Triggering a single additional spike in a cerebral cortical neuron
was recently demonstrated to cause a cascade of extra spikes in the
network that is likely to rapidly decorrelate the network's microstate.
The mechanisms involved in this extreme sensitivity of cortical networks
are currently not well understood. Here, we show in a minimal model
of cortical circuit dynamics that exponential state separation after
single spike and even single synapse perturbations coexists with dynamical
stability to infinitesimal state perturbations. We propose a unifying
picture of exponentially separating flux tubes enclosing unique stable
trajectories composing the networks' state spaces.
\end{abstract}
\maketitle
Understanding the dynamical characteristics of cerebral cortex networks
is fundamental for the understanding of sensory information processing
in the brain. Bottom-up investigations of different generic neuronal
network models have led to a variety of results ranging from stable
\cite{key-Zillmer,key-Jahnke,key-stable}
to chaotic dynamics \cite{key-VreeswijkSompolinsky,key-chaos,key-Latham}.
In top-down attempts to construct classification and discrimination
systems with such networks, the 'edge of chaos' was proposed to be
computationally optimal \cite{key-edge}. Near this transition between
ordered and chaotic dynamics, a network can combine the fading memory
and the separation property, both of which are important for the efficacy
of computing applications \cite{key-reservoir}. While fading memory
(information about perturbations of the microstate die out over time)
is achieved by a stable dynamics, the separation property (distinguishable
inputs lead to significantly different macrostates) is best supported
by a chaotic dynamics.

Widely used in reservoir computing \cite{key-reservoir} and one of
the most simple models of cortical circuits are networks of randomly
coupled inhibitory leaky integrate and fire (LIF) neurons \cite{key-Burkitt2006}.
These networks exhibit stable chaos, characterized by stable dynamics
with respect to infinitesimal perturbations despite an irregular network
activity \cite{key-Zillmer,key-Jahnke}.  They thus exhibit fading
memory. Whether and how such networks realize the separation property
is however unclear. 

Motivated by the recent observation that real cortical networks are
highly sensitive to single spike perturbations \cite{key-Latham},
we examine in this letter how single spike and single synapse perturbations
evolve in a formally stable model of generic cortical circuits. We
show that random networks of inhibitory LIF neurons exhibit negative
definite Lyapunov spectra, confirming the existence of stable chaos.
The Lyapunov spectra are invariant to the network size, indicating
that stable dynamics is representative for large networks, extensive
and preserved in the thermodynamic limit. Remarkably, in the limit
of large connectivity, perturbations decay as fast as in uncoupled
neurons. Single spike perturbations induce only minute firing rate
responses but surprisingly lead to exponential state separation causing
complete decoherence of the networks' microstates within milliseconds.
By examining the transition from unstable dynamics to stable dynamics
for arbitrary perturbation size, we derive a picture of tangled flux
tubes composing the networks' phase space. These flux tubes form reservoirs
of stability enclosing unique stable trajectories, whereas adjacent
trajectories separate exponentially fast. In the thermodynamic limit
the flux tubes become vanishingly small, implying that even in the
limit of infinitesimal weak perturbations the dynamics would be unstable.
This contradicts the prediction from the Lyapunov spectrum analysis
and reveals that characterizing the dynamics of such networks qualitatively
depends on the order in which the weak perturbation limit and the
thermodynamic limit are taken.

We studied large sparse networks of $N$ LIF neurons arranged on directed
Erd\"os-R\'enyi random graphs of mean indegree $K$. The neurons'
membrane potentials $V_{i}\in(-\infty,V_{\textrm{T}})$ with $i=1\dots N$
satisfy
\begin{equation}
\tau_{m}\dot{V}_{i}=-V_{i}+I_{i}(t)
\label{eq:LIF}
\end{equation}
between spike events. When $V_{i}$ reaches the threshold $V_{T}\equiv1$,
neuron $i$ emits a spike and $V_{i}$ is reset to $V_{R}\equiv0$.
The membrane time constant is denoted $\tau_{m}$. The synaptic input
currents are
\begin{equation}
I_{i}(t)=\sqrt{K}I_{0}-\frac{J_{0}}{\sqrt{K}}\tau_{m}\sum_{j\in\textrm{pre}(i)}
\sum_{s}\delta(t-t_{j}^{(s)}),
\label{eq:syncurrent}
\end{equation}
composed of constant excitatory external currents $\sqrt{K}I_{0}$
and inhibitory nondelayed $\delta$ pulses of strength $-J_{0}/\sqrt{K}$,
received at the spike times $t_{j}^{(s)}$ of the presynaptic neurons
$j\in\textrm{pre}(i)$. The external currents $I_{0}$ were chosen
to obtain a target network-averaged firing rate $\bar{\nu}$.

Equivalent to the voltage representation, Eq.~\eqref{eq:LIF}, is
a phase representation in which each neuron is described by a phase
$\phi_{i}\in[-\infty,1]$, obtained through $\phi_{i}=-\ln\bm{(}(V_{i}-\sqrt{K}I_{0})/(V_{R}-\sqrt{K}I_{0})\bm{)}/T_{i}^{\textrm{free}}$
(where $T_{i}^{\textrm{free}}=-\ln\bm{(}(V_{T}-\sqrt{K}I_{0})/(V_{R}-\sqrt{K}I_{0})\bm{)}$
is the interspike interval of an isolated neuron), with a constant
phase velocity and the phase transition curve $U(\phi_{i})=-\ln\bm{(}\exp(-\phi_{i}T_{i}^{\textrm{free}})+J_{0}/(KI_{0})\bm{)}/T_{i}^{\textrm{free}}$
describing the phase updates at spike reception. The neurons' phases
thus evolve, from time $t_{s-1}$ after the last spike in the network
through the next spike time $t_{s}$, at which neuron $j^{*}$ fires,
according to the map
\begin{equation}
\phi_{i}(t_{s})=
\begin{cases}
\phi_{i}(t_{s-1})+(t_{s}-t_{s-1})/T_{i}^{\textrm{free}} & 
\mathrm{for}\, i\neq i^{*}\\
U\bm{(}\phi_{i}(t_{s-1})+(t_{s}-t_{s-1})/T_{i}^{\textrm{free}}\bm{)} &
\mathrm{for}\, i=i^{*}.
\end{cases}
\label{eq:phasemap}
\end{equation}
The neurons postsynaptic to the spiking neuron $j^{*}$ are $i^{*}\in\mathrm{post}\,(j^{*})$.
We used the exact phase map \eqref{eq:phasemap} for numerically exact
event-based simulations and to analytically calculate the single spike
Jacobian $\mathbf{D}(t_{s})=\frac{\partial\vec{\phi}(t_{s})}{\partial\vec{\phi}(t_{s-1})}$:
\begin{equation}
D_{ij}(t_{s})=
\begin{cases}
d_{i^{*}}(t_{s}) & \mathrm{for}\, i=j=i^{*}\\
1-d_{i^{*}}(t_{s}) & \mathrm{for}\, i=i^{*},\, j=j^{*}\\
\delta_{ij} & \mathrm{otherwise}.
\end{cases}
\label{eq:jacobian}
\end{equation}
This matrix depends on the spiking neuron $j^{*}$ and on the phases
of the spike receiving neurons $i^{*}$ through the derivative of
the phase transition curve $d_{i^{*}}(t_{s})=\partial_{\phi}U\bm{(}\phi_{i^{*}}(t_{s}^{-})\bm{)}$
evaluated at time $t_{s}^{-}$ just before spike reception \cite{key-supplement}.
Describing the evolution of infinitesimal phase perturbations, the
single spike Jacobians \eqref{eq:jacobian} were used for numerically
exact calculations of all Lyapunov exponents $\lambda_{1}\ge\dots\ge\lambda_{N}$
in a standard reorthogonalization procedure \cite{key-Benettin1980}. 

\begin{figure}
\includegraphics[width=1\columnwidth]{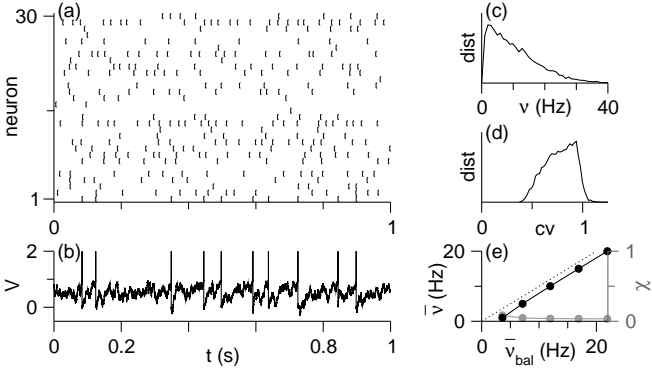}

\caption{\label{fig:characteristics}The balanced state in inhibitory LIF networks:
(a) Asynchronous irregular spike pattern of 30 randomly chosen neurons,
(b) fluctuating voltage trace of one neuron (voltage increased to
$V=2$ at spikes), (c),(d) broad distributions of individual neurons'
firing rates $\nu$ and coefficients of variation $\mathrm{cv}$,
(e) network-averaged firing rate $\bar{\nu}$ and synchrony measure
$\chi$ versus predicted rate $\bar{\nu}_{\textrm{bal}}=I_{0}/(J_{0}\tau_{m})$
(dotted line: guide to the eye for $\bar{\nu}=\bar{\nu}_{\textrm{bal}}$,
$\chi=\frac{\mathrm{STD}([\phi_{i}])}{[\mathrm{STD}(\phi_{i})]}$
where $[\cdot]$ denotes population average), (parameters: $N=10\,000$,
$K=1000$, $\bar{\nu}=10\,\textrm{Hz}$, $J_{0}=1$, $\tau_{\textrm{m}}=10\,\textrm{ms}$).}

\end{figure}

As expected from the construction of the LIF networks, the dynamics
converged to a balanced state. Figure~\ref{fig:characteristics}
shows a representative spike pattern and voltage trace illustrating
the irregular and asynchronous firing and strong membrane potential
fluctuations. A second characteristic feature of balanced networks
is a substantial heterogeneity in the spike statistics across neurons,
indicated by broad distributions of coefficients of variation (cv)
and firing rates ($\nu$). Independent of model details, the network-averaged
firing rate $\bar{\nu}$ in the balanced state can be predicted as
$\bar{\nu}_{\textrm{bal}}\approx I_{0}/(J_{0}\tau_{\textrm{m}})$
\cite{key-supplement}. The good agreement of this prediction with
the numerically obtained firing rate confirms the dynamical balance
of excitation and inhibition in the studied networks.

\begin{figure}
\includegraphics[clip,width=1\columnwidth]{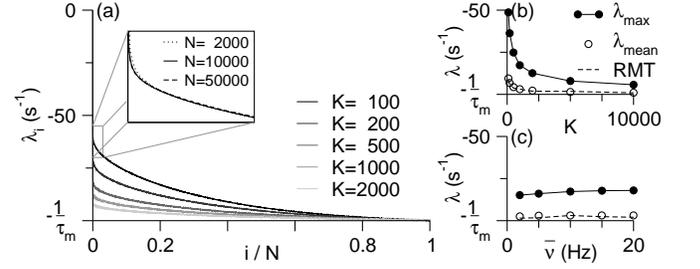}

\caption{\label{fig:lyapunovspectra}Stable dynamics with respect to infinitesimal
perturbations: (a) Spectrum of Lyapunov exponents $\{\lambda_{i}\}$
of networks of $N=10\,000$ LIF neurons for different connectivities
$K$, inset: close-up of spectra for $K=100$ and different network
sizes $N$, (b),(c) largest Lyapunov exponent $\lambda_{2}=\lambda_{\textrm{max}}$
and mean Lyapunov exponent $\lambda_{\textrm{mean}}=\frac{1}{N}\sum_{i=1}^{N}\lambda_{i}$
versus connectivity $K$ and average firing rate $\bar{\nu}$ (dashed
lines: random matrix theory for $\lambda_{\textrm{mean}}$ \cite{key-supplement}),
(parameters: $N=100\,000$, $K=1000$, $\bar{\nu}=10\,\textrm{Hz}$,
$J_{0}=1$, $\tau_{\textrm{m}}=10\,\textrm{ms}$; averages of $10$
initial conditions).}
\end{figure}

Although the voltage trajectory of each neuron and the network state
were very irregular, the collective dynamics of the networks was apparently
completely stable (Fig.~\ref{fig:lyapunovspectra}). For all firing
rates, coupling strengths and connection probabilities, the whole
spectrum of Lyapunov exponents (disregarding the zero exponent for
perturbations tangential to the trajectory) was negative, confirming
the occurrence of so-called stable chaos in LIF networks
\cite{key-Zillmer,key-Jahnke}. The invariance of the Lyapunov spectra to the
network size $N$, to our knowledge, for the first time demonstrates that this
type of dynamics is extensive. With increasing connectivity $K$ all Lyapunov
exponents approached a constant $\lambda_{i}\approx-1/\tau_{m}$. This is deduced
from the mean Lyapunov exponent given by
$\lambda_{\textrm{mean}}\approx-1/\tau_{m}+(V_{T}-\langle
V\rangle)/(\sqrt{K}I_{0})+\mathcal{O}(1/K)$
in random matrix approximation and the numerical observation that
the largest exponent approached $\lambda_{\mathrm{mean}}$ in the
large $K$-limit (\cite{key-supplement} and Fig.~~\ref{fig:lyapunovspectra}(b)).
These results suggest that in the thermodynamic limit arbitrary weak
perturbations decay exponentially on the single neuron membrane time
constant. As will become clear in the following, however, this issue
is quite delicate.

\begin{figure}
\includegraphics[clip,width=1\columnwidth]{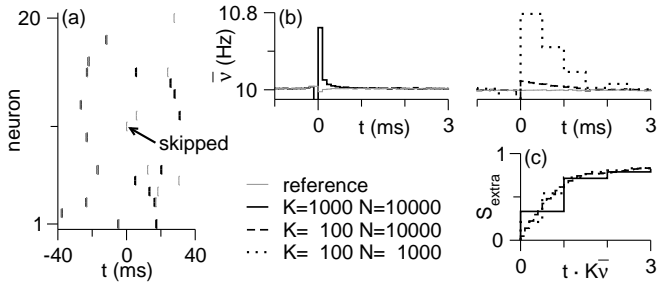}

\caption{\label{fig:rateresponse}Weak firing rate response after single spike
failure: (a) Sample spike pattern of 20 randomly chosen neurons (gray:
reference trajectory, black: single spike skipped at $t=0$), (b)
network-averaged firing rate of reference trajectory $\bar{\nu}$
and in response to skipped spike $\tilde{\nu}$ versus time for different
connectivities $K$ and network sizes $N$, (c) number of extra spikes
$S_{\textrm{extra}}=N\int(\tilde{\nu}-\bar{\nu})\textrm{d}t$ in the
entire network versus time (rescaled with average input rate $K\bar{\nu}$),
(parameters: $\bar{\nu}=10\,\textrm{Hz}$, $J_{0}=1$, $\tau_{\textrm{m}}=10\,\textrm{ms}$;
averages of $100$ initial conditions with $10\,000$ calculations
each).}
\end{figure}

Experimentally realizable and well-controlled state perturbations
to the dynamics of cortical networks are the addition or suppression
of individual spikes \cite{key-Latham,key-Brecht}. Such minimalistic
neurostimulation can elicit complex behavioral responses \cite{key-Brecht}
and can trigger a measurable rate response in intact cortical networks
\cite{key-Latham}. We therefore examined how such single spike perturbations
affected the collective dynamics of our networks. Here, the simplest
single spike perturbation is the suppression of a single spike. Figure~\ref{fig:rateresponse}
illustrates the firing rate response if one spike is skipped at $t=0$.
The missing inhibition immediately triggered additional spikes in
the $K$ postsynaptic neurons such that the network-averaged firing
rate increased abruptly by $\delta\bar{\nu}\sim K\bar{\nu}/N$. Since
the induced extra spikes inhibited further neurons in the network,
the overshoot in the firing rate quickly settled back to the stationary
state within a time of order $\delta t\sim1/(K\bar{\nu})$. The overall
number of additional spikes in the networks therefore was $N\delta\bar{\nu}\delta t\approx1$
and the one skipped spike was immediately compensated by a single
extra spike \cite{key-supplement}.

\begin{figure}
\includegraphics[clip,width=1\columnwidth]{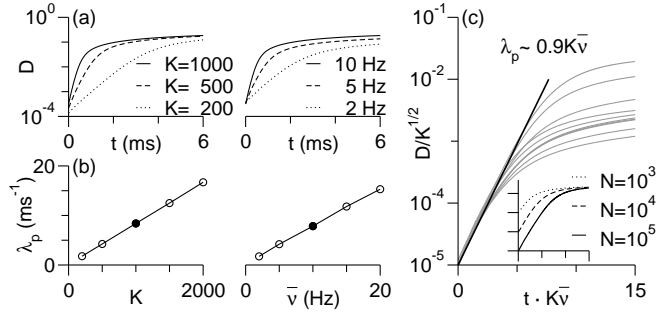}

\caption{\label{fig:distancespikeskipped}Sensitivity to single spike failures:
(a) Distance $D$ between trajectory after spike failure and reference
trajectory versus time in log-lin plots for different connectivities
$K$ and average firing rates $\bar{\nu}$, (b) pseudo Lyapunov exponent
$\lambda_{p}$ from exponential fits $D\sim\exp(\lambda_{p}t)$ before
reaching saturation versus connectivity $K$ and average firing rate
$\bar{\nu}$, (c) distance-evolution of all parameter sets (rescaled
with approximate perturbation strength $KJ_{0}/\sqrt{K}$) versus
time (rescaled with average input rate $K\bar{\nu}$) collapse to
characteristic exponential state separation with rate $\lambda_{p}\sim0.9K\bar{\nu}$
(inset: different network sizes $N$ for $K=100$), (parameters: $N=100\,000$,
$K=1000$, $\bar{\nu}=10\,\textrm{Hz}$, $J_{0}=1$, $\tau_{\textrm{m}}=10\,\textrm{ms}$;
averages of $10$ initial conditions with $100$ calculations each).}
\end{figure}

Even though the failure of one individual spike resulted in very weak
and brief firing rate responses, it nevertheless induced rapid state
decoherence. Figure \ref{fig:distancespikeskipped} displays the distance
$D(t)=\frac{1}{N}\sum_{i}|\tilde{\phi}_{i}(t)-\phi_{i}(t)|$ between
the perturbed trajectory (spike failure at $t=0$) and the reference
trajectory. After the spike failure, all trajectories separated exponentially
fast at a surprisingly high rate. Because this exponential separation
of nearby trajectories is reminiscent of deterministic chaos, we call
its separation rate the pseudo Lyapunov exponent $\lambda_{p}$. The
pseudo Lyapunov exponent was network size invariant, but showed a
completely different behavior compared to the classical Lyapunov exponents.
With increasing connectivity, it appears to diverge linearly $\lambda_{p}\sim K\bar{\nu}$.
It is thus expected to grow to infinity in the high connectivity limit,
reminiscent of binary neuron networks exhibiting an infinite Lyapunov
exponent in the thermodynamic limit \cite{key-VreeswijkSompolinsky}. 

\begin{figure}
\includegraphics[width=1\columnwidth]{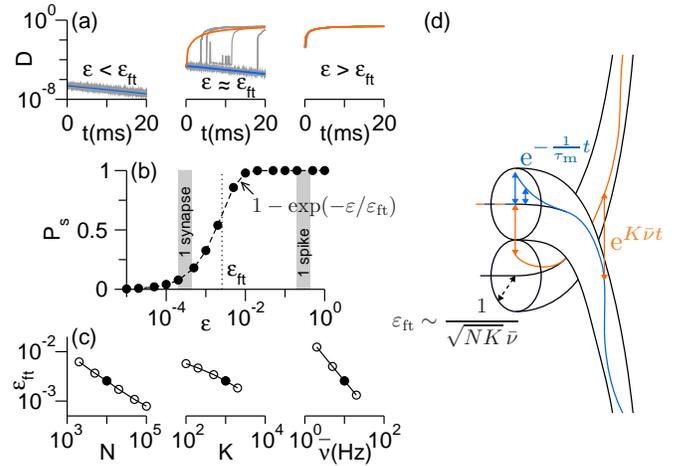}

\caption{\label{fig:distancefinitesize}(Color online) Sensitivity to finite-size
perturbations: (a) Distance $D$ between perturbed and reference trajectory
measured at spike times of reference trajectory (projecting out possible
time-shifts) for perturbations of strengths $\varepsilon=0.00002,\,0.002,\,0.2$
in log-lin plots (gray lines: 20 examples for initial perturbations
of same size pointing in different random directions perpendicular
to trajectory, color lines: averages of exponentially separating/converging
cases), (b) probability $P_{s}$ of exponential state separation versus
perturbation strength $\varepsilon$ in lin-log plot (dashed line:
fit to $P_{s}(\varepsilon)=1-\exp(-\varepsilon/\varepsilon_{\textrm{ft}})$,
dotted line: characteristic perturbation size $\varepsilon_{\textrm{ft}}$
separating stable from unstable dynamics, shaded areas: strengths
corresponding to single synapse and single spike failures), (c) characteristic
perturbation size $\varepsilon_{\textrm{ft}}$ versus network size
$N,$ connectivity $K$ and average firing rate $\bar{\nu}$ in log-log
plots, (d) symbolic picture of stable flux tubes with radius $\varepsilon_{\textrm{ft}}$
(stable dynamics inside flux tube but exponential separation of adjacent
flux tubes), (parameters: $N=10\,000$, $K=1\,000$, $\bar{\nu}=10\,\textrm{Hz}$,
$J_{0}=1$, $\tau_{\textrm{m}}=10\,\textrm{ms}$; averages of 10 initial
conditions with $100$ calculations and $100$ random directions each).}
\end{figure}

In the same balanced LIF networks, we thus find stable dynamics in
response to infinitesimal perturbations and unstable dynamics in response
to single spike failures. To further analyze the transition between
these completely opposite behaviors, we applied finite perturbations
of variable size perpendicular to the state trajectory (Fig.~\ref{fig:distancefinitesize}).
Depending on the perturbation strength $\varepsilon$ and direction
$\overrightarrow{\delta\phi}$ (with $\sum_{i}\delta\phi_{i}^{2}=1$),
the perturbed trajectory either converged back to the reference trajectory
or diverged exponentially fast. The probability $P_{s}(\varepsilon)$
that a perturbation of strength $\varepsilon$ induced exponential
state separation was very well-fitted by
$P_{s}(\varepsilon)=1-\exp(-\varepsilon/\varepsilon_{\textrm{ft}})$.
Hence, $\varepsilon_{\textrm{ft}}$ is a characteristic phase space
distance separating stable from unstable dynamics. Intriguingly, this
distance decreased as $\varepsilon_{\textrm{ft}}\sim1/(\sqrt{KN}\bar{\nu})$.
For large $K$ and $N$ the dynamics in the thermodynamic limit ($N\to\infty$)
would be unstable even to infinitesimal perturbations ($\varepsilon\to0$).
Contrary, the analysis of the Lyapunov spectra has shown that taking
the limit $\varepsilon\to0$ first and then $N\to\infty$ yields stable
dynamics. Thus, the order of the limits appears crucial in defining
the dynamical nature of balanced LIF networks.

The evolution of finite perturbations suggests a picture of stable
flux tubes around unique trajectories (Fig.~\ref{fig:distancefinitesize}(d)).
Perturbations within these flux tubes decayed exponentially, whereas
perturbations greater than the typical flux tube radius $\varepsilon_{\textrm{ft}}\sim1/(\sqrt{KN}\bar{\nu})$
induced exponential state separation. Single synaptic failures correspond
to small perturbations of size $\varepsilon_{\textrm{syn}}\approx J_{0}/\sqrt{KN}$
and therefore had a $N$ and $K$ independent probability of inducing
exponential state separation. This probability increased linearly
with the average firing rate $\bar{\nu}$ \cite{key-supplement}.

Summarizing, our analysis revealed the co-occurence of dynamical stability
to infinitesimal state perturbations and sensitive dependence on single
spike and even single synapse perturbations in the dynamics of networks
of inhibitory LIF neurons in the balanced state. They exhibit a negative
definite extensive Lyapunov spectrum that at first sight suggests
a well-defined thermodynamic limit of the network dynamics characterized
by stable chaos as previously proposed \cite{key-Zillmer,key-Jahnke}.
In this dynamics, single spike failures induce extremely weak firing
rate responses that become basically negligible for large networks.
Nevertheless, such single spike perturbations typically put the network
state on a very different dynamical path that diverges exponentially
from the original one. The rate of exponential state separation was
quantified with the so called pseudo Lyapunov exponent $\lambda_{p}$.
The scaling of $\lambda_{p}\sim K\bar{\nu}$ implies extremely rapid,
practically instantaneous, decorrelation of network microstates. Our
results suggest that the seemingly paradoxical coexistence of local
stability and exponential state separation reflects the partitioning
of the networks' phase space into a tangle of flux tubes. States within
a flux tube are attracted to a unique, dynamically stable trajectory.
Different flux tubes, however, separate exponentially fast. The decreasing
flux tube radius in the large system limit suggests that an unstable
dynamics dominates the thermodynamic limit. The resulting sensitivity
to initial conditions is described by the rate of flux tube separation,
the pseudo Lyapunov exponent, that showed no sign of saturation. These
findings suggest that the previously reported infinite Lyapunov exponent
on the one hand \cite{key-VreeswijkSompolinsky} and local stability
on the other hand \cite{key-Zillmer,key-Jahnke} resulted from the
order in which the weak perturbation limit and the thermodynamic limit
were taken.

For finite networks, the phase space structure revealed here may provide
a basis for insensitivity to small perturbations (e.g. noise or variations
in external inputs) and strong sensitivity to larger perturbations.
In the context of reservoir computing, the flux tube radius defines
a border between the fading property (variations of initial conditions
smaller $\varepsilon_{\mathrm{ft}}$ die out exponentially) and the
separation property (input variations larger $\varepsilon_{\mathrm{ft}}$
cause exponentially separating trajectories). Applications of LIF
neuron networks in reservoir computing may thus strongly benefit if
the flux tube structure of the network phase space is taken into account.
Our results of a very high pseudo Lyapunov exponent also reveal that
the notion of an 'edge of chaos' is not applicable in these networks.

We thank E.~Bodenschatz, T.~Geisel, S.~Jahnke, J.~Jost, P.~E.~Latham,
R.~M.~Memmesheimer, H.~Sompolinsky, M.~Timme and C.~van Vreeswijk
for fruitful discussions. This work was supported by BMBF (01GQ07113,
01GQ0811), GIF (906-17.1/2006), DFG (SFB 889) and the Max Planck Society.

\end{document}